\documentclass[epj]{svjour}
\usepackage{graphics,epsfig}
\usepackage{amssymb}
\usepackage{amsmath}
\usepackage{color}

\newcommand{\dd}{{\rm d}}

\begin{document}
\title{Links between nonlinear dynamics and statistical mechanics in a
simple one-dimensional model}
\author{
Hicham Qasmi, Julien Barr{\'e} \and Thierry
Dauxois\thanks{E-mail:Thierry.Dauxois@ens-lyon.fr}
}
\offprints{Thierry Dauxois} \institute{Laboratoire de Physique,
UMR-CNRS 5672, ENS Lyon, 46 All\'{e}e d'Italie, 69364 Lyon
C\'{e}dex 07, France}
\date{Received: date / Revised version: \today}
\abstract{We consider the links between nonlinear dynamics and
  thermodynamics in the framework of a simple nonlinear model for DNA.
  Two analyses of the phase transition, either with the transfer
  integral approach or by considering the instability of a nonlinear
  particular solution, are discussed. Conversely, the computation of
  the largest Lyapunov exponent is obtained within a thermodynamic
  treatment. Differences with the Peyrard-Bishop model are also
  discussed.
Lyapunov exponents
} 
\authorrunning{Qasmi, Barr{\'e} and  Dauxois}
\titlerunning{Links between nonlinear dynamics and statistical mechanics} \maketitle

\section{Introduction}

The main goal of the paper is to study the links between the
microscopic dynamics and macroscopic thermodynamical properties in
a very simplified model for DNA. Both aspects are usually not
studied simultaneously;  in the literature, the main goal is
often, either to consider  dynamical aspects of coherent
structures (solitons for example) in a system at zero temperature,
or to derive thermodynamical properties without really considering
the consequences of the existence of these coherent structures.
Here, on the contrary, we will put the emphasis on the link and
show that both approaches give important insights to the
description of the physical properties.

Two aspects will be of particular importance and we would like to
shed light on them already in the introduction. A recent
method~\cite{dauxois_peyrard}, which showed how the stability of a
nonlinear solution of the dynamical equations exhibits an
interesting approach to describe a phase transition, will be
applied in this new model: it will reveal how the small amplitude
dynamics needs a careful treatment to adequately describe the
thermodynamics. Second, we will describe the relationships between
the maximum Lyapunov exponent which characterizes the dynamics and
the thermodynamics. Using a geometric method~\cite{pettini}, we
will explicitly compute the evolution of this dynamical quantity
as a function of the temperature: consequences of the phase
transition will therefore be explicit. From these
calculations, we predict features absent in the Peyrard-Bishop model,
although both models are very similar: a non monotonic
behavior of the maximum Lyapunov exponent in the low temperature
phase, and a jump at the critical temperature.

In Sec.~\ref{model} we  present briefly the model. In the
following  Sec.~\ref{thermodynamics}, we  show how one can derive
its thermodynamical properties. Then, the emphasis is put in
Sec.~\ref{domainwall} to a special particular solution and on its use
to explain the thermodynamical properties. Finally, using the
powerful geometric method introduced recently to compute the
largest Lyapunov exponent, we discuss in Sec.~\ref{lyapunov} the
link between this dynamical quantity and the phase transition, a
thermodynamic concept.

\section{The Model}
\label{model}

A very simplified model has been proposed in 1989 by M. Peyrard and A.
R. Bishop~\cite{peyrardbishop} to describe DNA denaturation. This
biological phenomenon leads to the breaking of the H-bonds linking
both strands of DNA. This process, which appears when either the
temperature increases or when the pH of the surrounding solvent is
modified, was previously successfully described by Ising
models~\cite{wartellbenight}. However, the dynamical properties of the
phenomenon were not captured by these static descriptions. This is
why, keeping the principle of simplicity, Peyrard and Bishop described
this highly complicated biomolecule by two chains of particles coupled
by nonlinear springs. This step toward a more complex system, since
including the minimal dynamics, was unexpectedly more successful than
first understood and have lead to many studies (See
Ref.~\cite{peyrard} for a review). A large part of the results were
interesting from the physical point of view: one may list in
particular studies of discrete breathers modes and energy localization
in systems involving nonlinear and discreteness
effects~\cite{dauxoispeyrardbishop,dauxoispeyrardprl}, existence of
phase transition in one-dimensional system~\cite{dauxois_peyrard}\dots
Furthermore, several recent results have emphasized
that this model could be successfully used to locate promotor
regions~\cite{Rasmussenbishop} for real DNA sequences. This unexpected
report and similar ones explain why so simple models are still
nowadays thought to be possible powerful tools to describe real DNA
dynamics.

\begin{figure}[!h]
 \begin{center}
 \epsfig{file=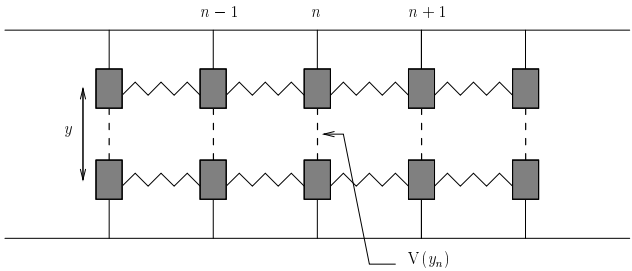,width=8.5 cm}
 \end{center}
 \caption{Schematic presentation of the DNA model.}\label{dessin}
 \end{figure}

Two linear chains of particles describing phenomenologically the
nucleotides describe the two different strands of DNA as
schematically represented in Fig.~\ref{dessin}. In the simplest
description, all particles of the same chain are harmonically
coupled whereas the interstrand interactions are restricted to
facing nucleotides; long-range interactions are neglected at this
level of description. It is important to understand that because
the goal is to describe denaturation dynamics, only transversal
degrees of freedom are taken into account. Defined with respect to
their equilibrium position, the displacement of the center of mass
of the $n^{{\rm th}}$ nucleotide are called $u_n$ (resp. $v_n$)
for the top (resp. bottom) chain.

Denoting $p_u$ the conjugated momentum to the spatial position $u$
and the number of nucleotides being $N$, the Hamiltonian model can
be written as
 \begin{eqnarray}
 H  = \sum_{n} \Biggl[&\displaystyle \frac{p_{u,n}^2}{2 m}  + \frac{K}{2} (u_n-u_{n-1})^2
 +&\nonumber\\
& \displaystyle\frac{p_{v,n}^2}{2 m} + \frac{K}{2} (v_n-v_{n-1})^2
+& V\left(\frac{u_n-v_n}{\sqrt{2}}\right)
 \Biggr].
 \end{eqnarray}
The interstrand potential $V$ describes the effective
interactions, i.e. in particular  hydrogen bonds between base
pairs but also the repulsion between phosphates. The canonical
transformation $x_n={(u_n+v_n)}/{\sqrt{2}}$ and
$y_n={(u_n-v_n)}/{\sqrt{2}}$ decouples simply both degrees of
freedom since  $H$ can be rewritten as $H =  H_x+H_y$ where
\begin{equation}
 H_x = \sum_n \left[{  \frac{p_{x,n}^2}{2 m}  + \frac{K}{2} (x_n-x_{n-1})^2 }\right]
 \end{equation}
 and
 \begin{equation}
H_y =  \sum_n \left[{ \frac{p_{y,n}^2}{2 m} + \frac{K}{2}
(y_n-y_{n-1})^2 + V\left(y_n \right) }\right].\label{defHamil}
\end{equation}

The dynamics and the thermodynamics of the first part, $H_x$,
which corresponds to a linear chain of harmonic oscillators, can
be easily computed. We will omit this part in the remaining of the
paper without loss of generality. The second part, $H_y$, on the
contrary needs further developments. For the sake of simplicity,
we will omit the $y$ index.

The system defined by this hamiltonian $H$ exhibits a second order
phase transition as shown in the next section. The low temperature
phase corresponds to states where the particles are located close
to their equilibrium position, the associated DNA being in the
native state: it must correspond to the bottom well of the
potential~$V$. On the contrary, in the high temperature phase, DNA
being denaturated, the link between facing nucleotides of both
strands are broken: consequently, associated particles of the
model must be located in a plateau of the potential, far from
their equilibrium positions since the force is vanishing. The
position in  this plateau  will be thermodynamically chosen
because of the entropy contribution to the free energy, important
only at high temperature.

This simple physical description guides therefore the appropriate
choice for the potential $V$. Nevertheless, the analytical
calculations that we will present now are possible for only a few
cases. The Morse potential  $V_m(y)=D(\exp(-a_my)-1)^2$ was the
first choice~\cite{peyrardbishop,dauxois_peyrard}. Here, we will
present another possible example~\cite{morse}:
 \begin{equation}\label{potmorsemorceaux}
V(y)= \left\{ \begin{array}{ll}
    -\displaystyle\frac{D}{\cosh^2 a y} &    \mbox{if}\ y\geq 0\\
    &\\
               +\infty &    \mbox{if }
                 y < 0.
  \end{array}
  \right.
\end{equation}
As we will see, the qualitative shapes are really close but the
differences of curvatures lead to several consequences. In
addition, the impossibility to reach negative values for the
variable $y$ gives interesting properties. One important
question concerns the influence of the details of the potential
$V$.  We will discuss in particular the following points:
\begin{itemize}
\item[i)] values of the critical temperature of the phase transition for both
potentials,
\item[ii)] consequences to the related characteristic lengths: order
parameter and correlation length,
\item[iii)]
largest Lyapunov exponent as a function of temperature.
\end{itemize}

\begin{figure}[!h]
\null\hskip 1.5truecm\epsfig{file=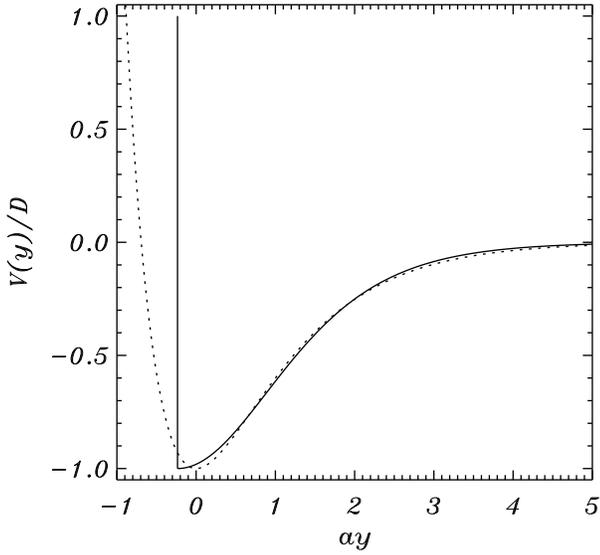,width=7cm}
\vskip 0.5truecm \caption{Comparison between
potential~(\ref{potmorsemorceaux}) (represented by the solid line)
with appropriate characteristics $(a,y_0)$ and the Morse potential
$V/D$ (dotted line). \label{fig:allure}}
\end{figure}

\section{Thermodynamics of the model}
\label{thermodynamics}
\subsection{The canonical partition function}

As it is well-known
nowadays~\cite{KrumhanslSchrieffer,dauxoispeyrardbook}, the
statistical mechanics of such a one-dimensional short-range
Hamiltonian can be exactly derived with the transfer integral
method (See appendix~\ref{transferinteg}). In the framework of
the continuum approximation, the solution relies on solving the
following Schr{\"o}dinger equation
\begin{equation}\label{equaschro}
 -\frac{1}{2 \beta^2 K}\frac{\dd^2\psi}{\dd y^2} -\frac{D}{\cosh^2 a y } \psi =  {\cal E}_k  \psi,
\end{equation}
if the lattice spacing between sites in the
$x$-direction is set to one.

 Defining the quantity
\begin{equation}
\eta=\frac{1}{4}\left[\sqrt{1 +8\frac{{T_c}^2}{T^2}}-1\right]
\end{equation}
where
\begin{equation} T_c=\frac{\sqrt{ K D}}{ak_B},\label{Tcexact}
\end{equation} this equation has $N_\eta =
E(\eta+1/2)$ localized states, with $E(.)$ denoting the integer part.
One notes that $T_c$ corresponds to the disappearance of the last
discrete state, and will be called the critical temperature.

The  $(N_\eta+1)$ localized states $\psi_k$ can be
expressed~\cite{morse} in terms of hypergeometric functions as
\begin{eqnarray}
 \psi_k(y)& = & \frac{{\cal N}_k}{\cosh(a y)^{b_{2 k+1}}} F\bigg(-2 k-1,2
 b_{2 k+1} + 2 k +2,\nonumber\\&&\hskip2.5truecm b_{2 k+1} +1; \frac{e^{- a y}}{e^{- a y}+e^{ a y}}\bigg),
\end{eqnarray}
where ${\cal N}_k$ is the normalization factor of the wave
function, $b_{n} =  2 \eta -n$ and finally
\begin{eqnarray}{\cal E}_k = -\frac{a^2}{2 \beta^2 K} \left({2\eta-2 k
-1}\right)^2,
\end{eqnarray}
the associated eigenvalues.

The ground state which is particularly useful in the remaining of
the paper can be simplified as
\begin{equation}
\psi_0(y)= \sqrt{ 2 a \frac{2 \eta -1}{B\left({2 \eta,
\frac{1}{2}}\right)}} \frac{\sinh a y}{\cosh^{2 \eta} a
y}\label{exprpsizero}
\end{equation} by introducing the Euler function
\begin{equation}
 B(x,y)=\frac{\Gamma(x+y)}{\Gamma(x)\Gamma(y)}=\int_{0}^{1} \dd t\, (1-t)^{x-1} t^{y-1}.
\end{equation}

\subsection{Discussion of the choice of the parameters set}

One of the goal of this work is
to make a detailed comparison between the Morse potential and
potential~(\ref{potmorsemorceaux}). The parameter set is obtained
either by considering  the effective physical interactions, or by
choosing the values to get the same melting temperature. Assuming
that the depth of the potential is known, 
both possibilities will define a relation between the width of the
potentials: $a_m^{-1}$ for the Morse one, and $a^{-1}$ for
potential~(\ref{potmorsemorceaux}).

The choice ${a_m} \approx 1.472\,a$ corresponds to the best fit. On the
contrary, as the transition temperature for the Morse potential
\cite{peyrardbishop,dauxois_these,dauxoispeyrardbook} is
${T_c^m}={T_c}\sqrt{8} $, the second choice $a_m=a\sqrt{8}$ would lead
to the same critical temperature.  Consequently, when the parameters
are fitted, the critical temperature differs by a factor two!

This unexpected disagreement reveals a hidden difference between
both potentials: the underlying reason is the important
contribution of negative positions $y$, in the Morse case. They
cannot be neglected even if the fast exponential increase was
thought to play the role of the impossibility of interpenetration.

The appropriate solution is to introduce an additional
parameter. The inverse width $a$ being given by the critical
temperature, one could adapt the minimum of the potential to minimize
the differences with the Morse potential. A
fit restricted to positive $y$-values over both variables
$({a_m},y_0)$ leads to an excellent agreement. This case
is presented in Fig.~\ref{fig:allure} by the solid line.

All results below correspond to the following set of parameters
$m=300$ u. a., $D=0.00094$ eV, $K=1.9$ eV.${\AA}^{-2}$ and $a=4.5$
${\AA}^{-1}$. The value of $T_c$ will be different from the Morse case,
but both potentials will be very similar as emphasized by
Fig.~\ref{fig:allure}. This choice will allow a precise study of the
negative $y$-values region and of the importance of potential
curvature for the Lyapunov exponent discussed in
section~\ref{lyapunov}.

\subsection{Characteristic lengths}

As usual, the thermodynamic properties of this system can be
characterized by an order parameter, and its fluctuations, in the
vicinity of the critical temperature $T_c$. Here, the appropriate
choice is the quantity
\begin{equation}
\ell =\left\langle \psi_0|x|\psi_0 \right\rangle
=\int_{0}^{\infty}\dd x \, x |{\psi_0}(x)|^2
,\label{sigmacontinu}
\end{equation}
which diverges for $T=T_c$. This clarifies the name critical
temperature which separates, the phase with a finite order
parameter (native state) from the phase with infinite order
parameter, representing the denaturated state.

\begin{figure}
\null\hskip 1.5truecm\epsfig{file=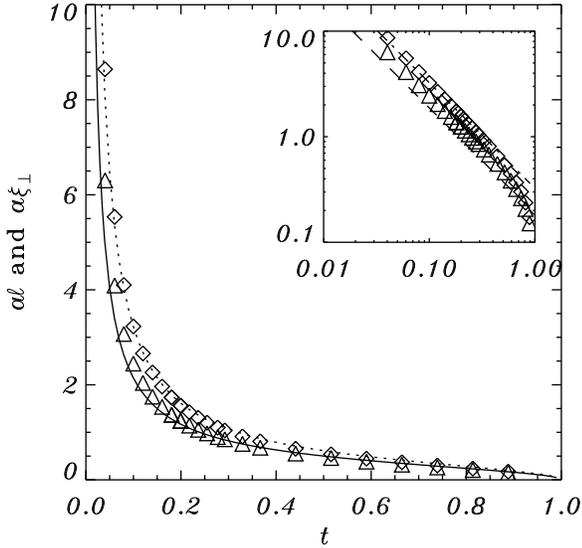,width=7cm}
\vskip 0.5truecm \caption{Order parameter $a \ell$ (solid line
and triangles) and correlation length $a \xi_\perp$ (dotted line
and diamonds) as a function of the reduced temperature
$t=1-T/T_c$. Lines corresponds to the transfer integral result
within the continuum approximation (Eqs.~(\ref{sigmacontinu})
and~(\ref{xicontinu})) whereas symbols correspond to the exact
numerical solution of the operator~(\ref{formuleA3}).  Presented
in the inset, the exact numerical data plotted with logarithmic
scales emphasize the critical behavior close to the critical
temperature. The dashed (resp. dash-dotted) line corresponds to
the asymptotic expression~(\ref{asympsig})
(resp.~(\ref{asympxi})).} \label{ordeparameter}
\end{figure}

The associated  critical exponent can even be determined by using
the asymptotic expression $\int_{0}^{\infty}\dd x\,
{x}/{\cosh^\alpha x}$ $ \stackrel{\alpha\to
0}{\sim}{1}/{\alpha^2}$. Using expression~(\ref{exprpsizero}) and
introducing the usual reduced temperature $t=1-{T}/{T_c}$, one
finally gets
\begin{equation}
{ \ell \stackrel{T\to T_c}{\sim} \frac{3}{16}\frac{1}
{a}\frac{1}{|{t}|} }.\label{asympsig}
\end{equation} In agreement with critical phenomena theory, we
obtain therefore the usual critical exponent $\beta=-1$, for this
second order phase transition. Fig.~\ref{ordeparameter} presents
the evolution of $\ell$ as a function of the temperature.
Formula~(\ref{sigmacontinu}), represented with a solid line,
agrees very well with the exact result obtained with the discrete
transfer operator~(\ref{formuleA3}).  In the inset, the
logarithmic plot emphasizes the critical behavior and confirms the
scaling exponent $\beta=-1$.

Interestingly it is also possible to determine the fluctuations of
the order parameter which can be computed from the following
expression
\begin{equation} {\xi_\perp^2}= \left\langle \psi_0|(x-\ell )^2|\psi_0
\right\rangle.\label{xicontinu}
\end{equation}
As above, the asymptotic expression  can be easily derived by
taking into account that $\int_{0}^{\infty}\dd
x\,{x^2}/{\cosh^\alpha x} \stackrel{\alpha\to 0}{\sim}
{2}/{\alpha^3}$. One thus obtains in the vicinity of the critical
temperature that
\begin{equation}{
\xi_\perp\stackrel{T\to T_c}{\sim} \frac{3\sqrt{3}}{16}\frac{1}{
a}\frac{1}{|{t}|} }.\label{asympxi}
\end{equation}
The critical exponent is therefore $\nu_\perp=1$.
Fig.~\ref{ordeparameter} shows that the order
parameter and its  fluctuations are of the same order in this
temperature regime. The inset confirms again the critical behavior
and the scaling exponent $\nu=1$.

It is important to stress that all above analytical results are
confirmed by the numerical but exact solution of the transfer
integral operator~(\ref{formuleA3}), without relying on the
continuum approximation: see symbols in Fig~\ref{ordeparameter}.
As it is nowadays known~\cite{dauxois_peyrard,dauxoispeyrardbook},
the critical exponents are valid even in the region where
discreteness effects can not be neglected. To be more specific,
let us introduce the parameter
\begin{equation}\label{defdeR}
R=\frac{D a^2}{K},
\end{equation}
measuring the onsite potential with respect to the elastic coupling.
For small value of this parameter, the continuum approximation is
valid and results~(\ref{asympsig}) and~(\ref{asympxi}) confirmed. For
the set of parameters chosen in this work, $R$ is equal to $10^{-2}$,
and one get results in values slightly different from
Eqs.~(\ref{asympsig}) and (\ref{asympxi}) for $T_c$, $\ell$, and
$\xi_{\perp}$. However, as shown by Fig.~\ref{ordeparameter} the
differences are hardly distinguishable and moreover the critical
exponents are fully confirmed.

\section{Domain Wall}
\label{domainwall}

In the framework of this model, it is also interesting to discuss
a new method~\cite{dauxois_peyrard,dauxoispeyrardbook},
alternative to the usual thermodynamic one proposed in previous
section, to detect the phase transition from dynamical
considerations. This will illustrate again the importance of the
wall located at $y=0$.

 Equations of motion corresponding to
Hamilto\-nian~(\ref{defHamil}) are
\begin{equation}
m{\ddot y}_{n}=K (y_{n+1}  + y_{n-1} -2 y_{n}) - \frac{ \partial
V}{\partial y_{n} }
\end{equation}
or, in the continuum limit,
\begin{equation}
{\ddot y} = \frac{K}{m}\frac{ \partial ^{2 }y}{\partial x^{2 } }
-\frac{ 1}{m } \frac{ \partial V}{\partial y }. \label{EMcont}
\end{equation}
The uniform profile at the minimum of $V(y)$ is a static solution of
the infinite chain with free ends. Profiles verifying $\dd^2 y /\dd
x^2 = 0$ are approximate solutions only on the plateau of the potential,
since $\partial V/\partial y$ is close to zero for large $y$.

There exists also an exact, unbounded, domain-wall like
solution
\begin{equation}
y_{DW}^{\pm}(x) = \frac{1}{a}\mbox{Argsh}\, e^{\pm z} = \frac{1}{a} \ln\left [ e^{\pm z} +\sqrt{1+e^{\pm 2z}}\right] , \label{DWsol}
\end{equation}
where $z=\sqrt{2R}(x-x_0)$ and $x_{0}$ is an arbitrary constant.
Solution (\ref{DWsol}) represents a  configuration
which links the stable minimum to a particular member of the
metastable configurations, with a slope $\sqrt{2D/K}$. One can easily
checks that this corresponds to equal contributions to the elastic and
the on-site potential energy densities ($D$ per site).  Consequently,
the energy of the solution contains a term which is proportional to
the number of sites to the right of $x_{0}$ and if lattice sites are
numbered from $0$ to $N$, one has
\begin{equation}
E_{DW}^{+}=\left( N -{ x_{0}}\right)2D + {\cal O}(N^{0
})  \quad. \label{DWen}
\end{equation}
At zero temperature the profile (\ref{DWsol}) is consequently not
stable, and the wall spontaneously move to the right,``zip\-ping''
back the unbound portion of the double chain.  This instability
changes however under the influence of temperature.

At non zero temperatures, let us consider small deviations with
respect to (\ref{DWsol}), i.e.
\begin{equation}
y(x,t) = y_{DW}^{+}(x-x_{0}) + \sum_{j} \alpha _{j}
f_{j}(x-x_{0})e^{-i\omega _{j}t}
\end{equation}
where $|\alpha _{j}|\ll a^{-1}$.  The linearized eigenfunctions $f_{j}$
satisfy the  Schr{\"o}dinger-like equation
\begin{equation}
-\frac{\dd^{2 }f_{j}}{ \dd z^{2 }} +
\frac{1-2e^{2z}}{\left(1+e^{2z}\right)^2}f_{j} = \frac{ m\omega
_{j}^{2 }}{2KR }\, f_{j}. \label{lin}
\end{equation}
Eq. (\ref{lin}) has no bound states~\cite{morse}. There are
however scattering states: acoustic phonons oscillating on the
flat portion of the  potential with frequencies
\begin{equation}
 \omega_{ac}^2=\frac{2K}{m} (1-\cos q)  \label{softcontinuumphonon}
\end{equation}
are some of them.

 In the bottom of the well of the potential, let
us first forget the wall. Consequently scattering states would be
optical phonons with frequencies
\begin{equation}
 \omega_{opt}^2 = \frac{2Da^2}{m}+\frac{2K}{m} (1-\cos q).\label{hardcontinuumphonon}
\end{equation}

At finite temperatures, the domain wall would therefore be
accompanied by a phonon cloud contributing to the free energy as
\begin{equation}
F_{ph} =  k_{B}T x_{0}\int _{0}^{\pi}\frac{ \dd q}{\pi  } \ln
\frac{\omega_{opt} }{\omega_{ac} } + ... \label{phcl}
\end{equation}
where we omit terms independent of $x_{0}$. Introducing the
dispersion relations, we can evaluate~\cite{gradshteyn} the
integral in~(\ref{phcl}) using $\int_{0}^{\pi} \dd x \ln [1-\cos x
+ R]= {2\pi  } \ln [ {(\sqrt{R}+\sqrt{R+2} )}/2].$ We obtain thus
the total free energy (DW plus phonon cloud)
\begin{equation}
F = \left( k_{B}T\ln \left[ \frac{\sqrt{R}+\sqrt{R+2} }{\sqrt{2} }
\right] -2D \right){ x_{0}} + const . \label{FDW}
\end{equation}
This  result
describes in very simple terms why and when the phase transition
occurs. At temperatures lower than
\begin{equation}
T_{c}=\frac{ 2D}{k_{B}\ln\left[ \sqrt{R/2} + \sqrt{1+R/2}  \right]
},  \label{Tcdiscrete}
\end{equation}
the prefactor of $x_{0}$ in Eq.~(\ref{FDW}) is negative,
describing the DW's natural tendency towards high positive values
of $x_{0}$: it ``zips'' the system back to the bound
configuration. Conversely, at temperatures higher than $T_{c}$,
thermal stability is achieved and the DW ``opens up".

It should be noted that the value of $T_{c}$ predicted by the
above DW argument coincides exactly with the result obtained for
the Morse potential~\cite{dauxois_peyrard,dauxoispeyrardbook}. The
limiting behavior in the continuum approximation, i.e. in the
limit $R\ll 1$, leads for example to $T_c=2\sqrt{2KD}/(ak_B)$.
This result differs nevertheless from Eq.~(\ref{Tcexact}).

Expression~(\ref{Tcdiscrete}) is consequently {\em not} valid for
the sech$^2$-potential~(\ref{potmorsemorceaux}) of interest. The
underlying reason is the impossibility of usual phonons to take
place in the bottom of the well. The harmonic approximation of the
problem leads indeed to a {\em nonlinear} problem because of the
nonlinear condition $y>0$ introduced by the wall. Optical phonons
with frequencies~(\ref{hardcontinuumphonon}) are therefore totally
modified by the wall and cannot be computed. A way to
take into account 
this nonlinear condition would consist in selecting only the even modes 
of optical phonons~(\ref{hardcontinuumphonon}). However, it turns out to be 
unexact and unsufficient. This illustrates 
again that the presence of the wall strongly modifies the {\em
dynamics} of the system, which leads to important {\em
thermodynamic} consequences.

\section{Geometrical method to derive the largest Lyapunov exponent}
\label{lyapunov}

In the theory of dynamical systems, the concept of {\em Lyapunov
exponent} has also attracted a lot of
attention~\cite{manneville,lichtenberg} because it defines
unambiguously a sufficient condition for chaotic instability.
Unfortunately, except for very few systems, it is already an
extremely difficult task to derive analytically the expression of
the largest one, $\lambda_1$, as a function of the energy density.
As some promising results have been recently obtained to describe
some properties of high-dimensional dynamical
systems~\cite{livi,Constantoudis,ruffo,firpo,VallejosAntenodo,Kurchan}, by combining tools
developed in the framework of dynamical systems with concepts and
methods of equilibrium statistical mechanics, the idea that both
concepts could be related was proposed~\cite{pettini}.

\subsection{Riemannian geometry approach} \label{sec:riemman}

The main idea is that the  chaotic hypothesis is at the origin of
the validity of equilibrium statistical physics, and this fact
should be traced somehow in the dynamics and therefore in the
largest Lyapunov exponent. The method is based on a reformulation
of Hamiltonian dynamics in the language of Riemannian
geometry~\cite{pettini}: the trajectories are seen as geodesics of
a suitable Riemannian manifold. The chaotic properties of the
dynamics are then directly related to the curvature of the
manifold and its fluctuations. Indeed, negative curvatures tend to
separate initially close geodesics, and thus imply a positive
Lyapunov exponent; nevertheless chaos may also be induced by
positive curvatures, provided they are fluctuating, through a
parametric-like instability. To approximate the curvature felt
along a geodesic, the method uses a Gaussian statistical process.
The mean value of this process is given by $\kappa_0$ and its
variance by $\sigma_\kappa$, where $\kappa_0 $ and $\sigma_\kappa$
are the statistical average of the curvature and its fluctuations,
which can be computed by standard methods of statistical
mechanics.

Finally one ends up with the following expression of the largest
Lyapunov exponent~\cite{pettini}
\begin{equation}\label{exprlyapunov}
\lambda_1=\frac{1}{2}\left(\Lambda-\frac{4\kappa_0}{3\Lambda}\right)\end{equation}
where
\begin{equation}\Lambda=\left(\sigma_\kappa^2\tau+\sqrt{\left(\frac{4\kappa_0}{3}\right)^3
+\sigma_\kappa^4\tau^2}\right)^{1/3}.\label{biglambda}
\end{equation}
In this definition, $\tau$, the relevant time scale associated to the
stochastic process, is function of the two following timescales:
$\tau_1\simeq{\pi}/{2\sqrt{\kappa_0+\sigma_\kappa}}$ is the time
needed to cover the distance between two successive conjugate points
along the geodesics, whereas
$\tau_2\simeq{\sqrt{\kappa_0}}/{\sigma_\kappa}$ is related to the
local curvature fluctuations. The general rough physical estimate
${\tau}\simeq\left({1}/{\tau_1}+{1}/{\tau_2}\right)^{-1}$ completes
finally the analytical estimate of $\lambda_1$.  We continue now by
calculating the mean value of the curvature and its fluctuations as a
function of the energy density.

\subsection{Average curvature}

The curvature of the Riemannian manifold is directly
  given by the Laplacian of the potential. One needs therefore to
  compute the microcanonical average of the quantity
\begin{eqnarray}
\Delta V & = & 2 K N + 2 a^2 D \sum_k g(y_k)
\label{formcurvmicro}\end{eqnarray} and its corresponding
fluctuations, where
\begin{eqnarray}
g(y)& = &\frac{3}{\cosh^4 a y}-\frac{2}{\cosh^2 a y}.
\end{eqnarray}
We finally obtain 
\begin{eqnarray}\langle{\Delta V }\rangle_\mu =N\left(2 K  + 2 a^2 D \langle
g(y_k)\rangle_{\mu}\right). \label{formmicro}\end{eqnarray}

As, in the thermodynamic limit, ensemble equivalence ensures that
averages are equal in all statistical ensembles, we will compute
them in the canonical one since the transfer operator method has
been shown to be a powerful tool to compute thermodynamic
functions, especially if the continuum approximation is valid.

Calculation of Eq.~(\ref{formmicro}) relies on the computation
of $\langle{g(a y)}\rangle_\text{can}$, i.e. of terms such as
$\langle{{1}/{\cosh^{2 \alpha}a y}}\rangle_\text{can}$. Using the
transfer operator method in the continuum framework, we
immediately  find
\begin{equation}
\left\langle{\frac{1}{\cosh^{2 \alpha} a
y}}\right\rangle_\text{can}=\int_{0}^{\infty}\dd y\,
\frac{1}{\cosh^{2 \alpha}a y}\,|{\psi_0(y)}|^2.
\end{equation}
With expression~\cite{gradshteyn}
\begin{equation}
I(\alpha)=\int_{0}^{\infty} \frac{\dd x}{\cosh^{2\alpha}
x}=B\left({2 \alpha, \frac{1}{2}}\right),
\end{equation}
and  the mathematical formula $I(\alpha+1) = I(\alpha){2 \alpha}/
{(2 \alpha +1)}$, canonical averages can be simplified. Introducing
the parameter $R$ defined in Eq.~(\ref{defdeR}), one gets
\begin{equation} \kappa_0 =\frac{\langle{\Delta V }\rangle_\mu }{N}=  2 K\left({1  +4R
\frac{(2\eta-1)(2\eta-3)}{(4\eta+1)(4\eta+3)}}\right).\label{expresionkappa0}
\end{equation}
Above expression gives in particular the following limiting
behaviors
\begin{equation}
\lim_{T \to 0} \kappa_0=2 K+2D a^2
\end{equation}
and
\begin{equation}
\lim_{T \to T_c} \kappa_0=2K,
\end{equation} which coincide with asymptotic results for the
Morse potential~\cite{barre_dauxois}.

\begin{figure}[h]
  \epsfig{file=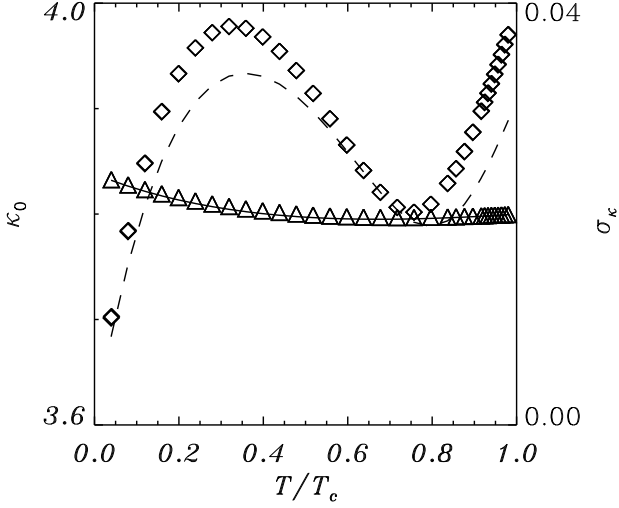,width=8.5cm} \caption{Average curvature
    $\kappa_0$ and canonical fluctuations $\sigma_\kappa$ versus temperature. The
    triangles (resp. diamonds) represent the exact numerical result
    for the curvature (resp. canonical fluctuations), obtained with
    the transfer integral formalism. The solid line corresponds to
    analytical formula~(\ref{expresionkappa0}) whereas the dashed line
    correspond to the microcanonical fluctuations~(\ref{correctmicro}).
    \label{fig:m_vs_s}}
\end{figure}

It is however important to notice that
expression~(\ref{expresionkappa0}) suggests that the mean curvature is
not always positive, its sign being tuned by the value of $R$.  As
negative curvatures enhance dynamical instability, this result may
have strong consequences on the largest Lyapunov exponent. A careful
study~\cite{rapporthicham}, shows that for values of $R$ larger than
$R_c=({31+3\sqrt{105}})/{8}$, expression~(\ref{expresionkappa0}) could
be negative in a given interval of temperatures. This result has to be
criticized since expression~(\ref{expresionkappa0}) has been derived
in the continuum approximation, and one expects important discreteness
effects for $R$ values as large as $R_c$. Numerical, but exact,
resolution of the transfer integral operator shows that the curvature
is actually positive for all $R$.  Fig.~\ref{fig:m_vs_s} shows the
curvature as a function of temperature for the parameter set chosen in
this work.

\subsection{Fluctuations of the curvature}

Contrary to the statistical averages such as the curvature~$\kappa_0$,
fluctuations are ensembles dependent.  This is why one computes first
fluctuations in the canonical ensemble, before using the
Lebowitz-Percus-Verlet formula~\cite{lebo} to get the microcanonical
fluctuations.

In the canonical ensemble, using Eq.~(\ref{formuleaciter}),
fluctuations of expression~(\ref{formcurvmicro}) are given by
\begin{eqnarray}
\frac{\langle{\delta^2\Delta V}\rangle}{4 a^4 D^2N}\!\!\!\!\!
\!\!\!\!&\!\!=\!\!&\!\!\!\!\frac{1}{N}
\sum_{i,j}\left[{\langle{g(y_i)g(y_j)}\rangle-\langle{g(y_i)}\rangle
\langle{g(y_j)}\rangle}\right] \\
             &=&\!\! {\sum_{l} \langle{g(y_N)g(y_{N-l})}\rangle-N\langle{g(y)}^2}\rangle \\
            &=& \!\!\sum_{l}\sum_{q=0}^{\infty} e^{-\beta l (\varepsilon_q-\varepsilon_0)}
    \left|\int\dd y g(y)\psi_q^\star(y)\psi_0(y) \right|^2 \nonumber\\
&&\hskip 3.4truecm - N \langle g(y)^2\rangle \\
             &=& \!\!\!\! \!\! \sum_{l}\sum_{q=1}^{\infty} e^{-\beta l (\varepsilon_q-\varepsilon_0)}
             \left|{\int{}{}\dd y g(y)\psi_q^\star(y)\psi_0(y)  }\right|^2  \\
             & \underset{ N\to \infty}{\simeq} &  \!\!\!\! \sum_{q=1}^{\infty}
\frac{1}{1-e^{-\beta  (\varepsilon_q-\varepsilon_0)}}
             \left|{\int{}{}\dd y\, g(y)\psi_q^\star(y)\psi_0(y)
}\right|^2.\label{formfluctfinal}
\end{eqnarray}
Above expression can be used to compute the canonical
fluctuations.

The microcanonical fluctuations will finally be recovered by using
the Lebovitz-Percus-Verlet formula~\cite{lebo}. For any quantity
$C$ with fluctuations  $\langle{\delta^2 C}\rangle$, both
fluctuations are related through the formula
\begin{equation}
\langle{\delta^2 C}\rangle_\mu = \langle{\delta^2
C}\rangle_\text{can}+\frac{\partial\langle{U}\rangle_\text{can}}{\partial\beta}^{-1}
\left(\frac{\partial\langle{C}\rangle_\text{can}}{\partial\beta}\right)^2,
\end{equation}
where $\langle{U}\rangle_\text{can}$ is the averaged energy of the
system.

The canonical partition function $Z$ being the product of a
kinetic part $Z_T$ and a configurational one $Z_c$, the averaged
energy is given by
\begin{equation}
\langle{U}\rangle_\text{can} = -\frac{\partial\ln
Z_T}{\partial\beta}-\frac{\partial\ln Z_c}{\partial\beta}.
\end{equation}
The first contribution is  as usual ${N}/({2\beta})$ whereas the
last one, can be simplified in the continuum approximation by
using the transfer integral method (See
Appendix~\ref{transferinteg} and in particular
Eq.~(\ref{formulaZconf})). Denoting $\varepsilon_0$ the ground
state of the associated Schr{\"o}dinger equation, one obtains $\ln
Z_c=-N\beta \varepsilon_0$ with
\begin{eqnarray}
 \varepsilon_0 &=& \frac{1}{2\beta}\ln\frac{\beta K}{2\pi}
-\frac{a^2}{2 K \beta^2 }(2\eta-1)^2  .
\end{eqnarray}
The averaged energy is therefore
\begin{equation}
\langle{U}\rangle_\text{can} = N\left[\frac{1}{2\beta} +
\frac{\partial(\beta \varepsilon_0)}{\partial\beta} \right].
\end{equation}
The final analytical expression for the microcanonical
fluctuations is not simple. Introducing quantities $\beta_c=1/(k_BT_c)$
and $\delta=T_c/T$, one gets \begin{eqnarray}
\frac{\langle{\delta^2\kappa_0}\rangle_\mu-\langle{\delta^2\kappa_0}\rangle_\text{can}}{18432 a^4 D^2 K\beta_c}
 &=& -   \delta^5
    \left({-16\delta^2+7\sqrt{1+8\delta^2}+5}\right)^2\nonumber
    \\ && / 
  (1+8\delta^2)^{\frac{3}{2}} /\left({2+\sqrt{1+8\delta^2}}\right)^4 \nonumber
    \\ && /  ( 16 K \beta_c \delta^3
      \sqrt{1+8\delta^2} -36 a^2 \delta^2 \nonumber
    \\ && +40 a^2 \delta^2 \sqrt{1+8\delta^2} + 2 K \beta_c \delta\nonumber
    \\ &&
   \! \! \! \! \! \! \!  \!  \! \! \sqrt{1+8\delta^2} -3 a^2 +5 a^2 \sqrt{1+8\delta^2}) .\label{correctmicro}
\end{eqnarray}
Above expressions~(\ref{formfluctfinal}) and~(\ref{correctmicro})
can be combined to compute the microcanonical fluctuations. This
is what has been performed to plot the fluctuations of the
curvature in Fig.~\ref{fig:m_vs_s}.

Close to the critical temperature $T_c$, the continuum part of the
transfer operator spectrum should be taken into account and an
explicit analytical calculation is possible in principle but
particularly tedious. On the contrary, one can simplify above
expression in the low temperature regime as shown in next section.

\subsection{Low temperature estimate}

In the low temperature regime by replacing the prefactor
$(1-\exp[-\beta (\varepsilon_q-\varepsilon_0)])^{-1}$ by 1 in
Eq.~(\ref{formfluctfinal}), one finally gets
\begin{eqnarray}
\langle{\delta^2\Delta V}\rangle_\text{can} & \simeq & 4N a^4 D^2 \sum_{q=1}^{\infty}  \left\langle \psi_q | g |\psi_0\right\rangle  \left\langle \psi_0 | g |\psi_q\right\rangle\\[2mm]
& = & 4N a^4 D^2\left[{ \left\langle \psi_0 | g^2
|\psi_0\right\rangle-  \left\langle \psi_0 | g
|\psi_0\right\rangle^2}\right].
\end{eqnarray}
Combining result \begin{equation}
 \left\langle \psi_0 |g^2|\psi_0\right\rangle=16\frac{2\eta(2\eta-1)(4\eta^2-6\eta+11)}
{(4\eta+1)(4\eta+3)(4\eta+5)(4\eta+7)}
\end{equation}
with formula~(\ref{expresionkappa0}), one ends up with
\begin{equation}
 \langle{\delta^2 g}\rangle_\text{can} = 48\frac{(2\eta-1)
(256\eta^3-184\eta+105)}{(4\eta+1)^2(4\eta+3)^2(4\eta+5)(4\eta+7)}.
\end{equation}

As the microcanonical correction~(\ref{correctmicro}) can be
neglected in this region, the microcanonical fluctuations of the
curvature are finally given in the low temperature regime by
\begin{eqnarray}
\sigma_\kappa^2 &=& \frac{\langle{\delta^2\Delta V}\rangle_\text{can} }{N}\\
&\simeq&
4D^2a^4\frac{48(2\eta-1)(256\eta^3-184\eta+105)}{(4\eta+1)^2(4\eta+3)^2(4\eta+5)(4\eta+7)}
 .\label{formfinalfluct}
\end{eqnarray}
This expression can be used in the low temperature region to get a
simpler expression for the largest Lyapunov exponent. One thus 
obtain that it increases quadratically with the temperature.

\subsection{Largest Lyapunov exponent}

We can now estimate the largest Lyapunov exponent $\lambda_1$ for
this high-dimensional system of $N$ coupled particles in the
external sech$^2$-potential~(\ref{potmorsemorceaux}).

As $\kappa_0$ is positive, the instability of trajectories is due
to fluctuations of curvatures as reported by Pettini, Casetti and
Cohen~\cite{pettini}. Analytical
expressions~(\ref{expresionkappa0}) for $\kappa_0$
and~(\ref{formfinalfluct}) for $\sigma_\kappa$ were the only
missing points in the estimate~(\ref{exprlyapunov}) for
$\lambda_1$. One can in addition notes that in the low temperature
region, where $\sigma_\kappa\ll\kappa_0$ as attested by
Fig.~\ref{fig:m_vs_s}, one gets the asymptotic expression
\begin{equation}
 \lambda_1 \simeq\frac{\pi}{8} \frac{\sigma_\kappa^2}{{\kappa_0}^\frac{3}{2}}.\label{aprroxexpr}
\end{equation}

\begin{figure}[h]
  \epsfig{file=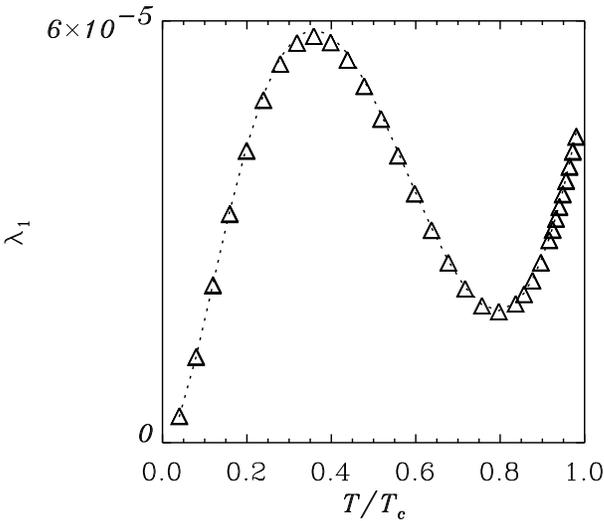,width=8.5cm} \caption{Largest Lyapunov
    exponent $\lambda_1$ versus ${T}/{T_c}$. The triangles correspond to the
    transfer integral result, whereas the dotted line correspond to
    the analytical expression (\ref{aprroxexpr}).}
\label{fig:lyapunov_m_vs_s}
\end{figure}

Fig. (\ref{fig:lyapunov_m_vs_s}) presents the temperature evolution of
the largest Lyapunov exponents. Two features should be emphasized in
comparison to what has been previously reported for the Morse
potential \cite{barre_dauxois}. One can notice a local maximum and a
local minimum of the Lyapunov exponent in the low temperature region.
More importantly, one has to realize that for temperatures larger than
the critical one $T_c$, particles are on the plateau of the potential;
the chain being equivalent to a linear chain, the largest Lyapunov
exponent $\lambda_1$ if of course zero. The Lyapunov exponent should
thus present a jump close to the critical temperature. It is important
to check these two strong predictions by considering careful
microcanonical numerical simulations with large systems. This would
provide a precise test of the geometrical method to calculate Lyapunov
exponents.

\section{Conclusion}

We have presented a new qualitative model for DNA denaturation
directly inspired by previous
works~\cite{peyrardbishop,dauxois_these,dauxois_peyrard}. Its
complete statistical mechanics was derived, as well as all features
related to the second order phase transition:
not only the critical temperature, but also the critical exponents
related to the order parameter and the transversal correlation
length.

We have in particular emphasized the important role of the negative
$y$-values for the Morse potential which were believed to be
unimportant~\cite{dauxois_these}. If critical exponents are of course
not affected, the critical temperature is strongly dependent on it.
Furthermore, using a geometric approach to estimate the largest
Lyapunov exponent, we have computed its evolution as a function of the
temperature. The results are unexpectedly qualitatively different from
those obtained with the Morse potential.

\acknowledgement This work has been partially supported by the
French Minist{\`e}re de la Recherche grant ACI jeune
chercheur-2001 N$^\circ$ 21-31, and the R{\'e}gion Rh{\^o}ne-Alpes
for the fellowship N$^\circ$ 01-009261-01.

\appendix
\section{Transfer integral method for the canonical partition function}
\label{transferinteg}

With periodic boundary conditions, $y_0=y_N$, the configurational
partition function of Hamiltonian~(\ref{defHamil}) can be written
as
\begin{equation}
Z_c = \int\prod_{i=0}^{N} \dd y_i\, e^{-\beta f(y_i,y_{i-1})}
\,\delta(y_0-y_N)\label{Zc}
\end{equation}
by introducing the symmetric function
\begin{eqnarray}
f(y_n,y_{n-1})  &=& -\frac{D}{2}\left[\frac{1}{\cosh^2 a y_n}
+\frac{1}{\cosh^2 a y_{n-1}}\right]\nonumber\\
&&+\frac{K}{2}(y_n-y_{n-1})^2. \end{eqnarray} Defining the
transfer operator $T$ as
\begin{eqnarray}
T[\phi] (y) = \int_{\mathbb{R}}\dd x\, \phi(x) e^{-\beta f(y,x)}
,\label{formuleA3}
\end{eqnarray}
its eigenvalues $\varepsilon_k$ and normalized eigenvectors
$\psi_k$ are given by $T[\psi_k]  =
\exp\left({-\beta\varepsilon_k}\right)\: \psi_k$.

Introducing the orthonormalization condition
\begin{eqnarray}
\delta(y-y_0)=\sum_{k}\psi_k^\star(y_0)\psi_k(y)\end{eqnarray} in
Eq.~(\ref{Zc}), one gets~\cite{dauxoispeyrardbook}
\begin{eqnarray}
Z_c =  \sum_{k} e^{-N\,\beta\,\epsilon_k}.
\end{eqnarray}
If the lowest eigenvalue $\varepsilon_0$ is discrete and located
in the gap below the continuum, one can simplify above expression
in the limit $N\to\infty$, so that
\begin{equation}
Z_c \simeq e^{-N\beta\varepsilon_0}.\label{formulaZconf}
\end{equation}

A similar method to compute the canonical average of any function
$h(y)$ leads to the following result
\begin{eqnarray}
\langle{h(y)}\rangle_\text{can} = \langle{h(y_N}\rangle_\text{can}
    & \underset{N\to\infty}{\simeq} & \int{}{} \dd y \,h(y)
    |\psi_0(y)|^2.\label{formuleaciter}
\end{eqnarray}

Above results are valid without approximations. However, applying
the continuum approximation, it is possible to go one step further
since there is a mapping~\cite{dauxois_these,dauxoispeyrardbook}
between the transfer integral operator and the following
Schr{\"o}dinger equation
\begin{equation}\label{equaschroappendix}
 -\frac{1}{2 \beta^2 K}\frac{\dd^2\psi}{\dd y^2} -\frac{D}{\cosh^2 a y }\psi =  {\cal E}_k  \psi  .
\end{equation}
As the spectrum of a quantum particle in the
potential~(\ref{potmorsemorceaux}) is known, one can derive the
analytical expression of~(\ref{formuleaciter}) within this
approximation.

\end{document}